\documentclass[aps,prb,twocolumn,showpacs,preprintnumbers,amsmath,amssymb,superscriptaddress,noshowkeys]{revtex4-1}
\usepackage{txfonts}
\usepackage[dvips]{graphicx}
\usepackage{bm}
\usepackage{color}
\usepackage{ulem} 
\def\rnum#1{\expandafter{\romannumeral #1}} 
\def\Rnum#1{\uppercase\expandafter{\romannumeral #1}}

\newfont{\bg}{cmr10 scaled\magstep4}

\newcommand{\bigzerou}{\smash{\lower1.8ex\hbox{\bg 0}}}
\begin{document}
\title{Classification of Weyl point trajectories in multi-terminal Josephson junctions}
\author{Kento Takemura}
\affiliation{Department of Material Engineering Science,
Graduate School of Engineering Science, Osaka University,
1-3 Machikaneyama, Toyonaka, Osaka 560-8531, Japan}
\author{Tomohiro Yokoyama}
\email[E-mail me at: ]{tomohiro.yokoyama@mp.es.osaka-u.ac.jp}
\affiliation{Department of Material Engineering Science,
Graduate School of Engineering Science, Osaka University,
1-3 Machikaneyama, Toyonaka, Osaka 560-8531, Japan}

\date{\today}

\begin{abstract}
Topological protection is an attractive signature in both fundamental and applied researches because it provides an exotic and robust state.
Multi-terminal Josephson junctions have recently been studied extensively owing to the emergence of topologically protected Weyl points
without the need for topological materials.
In this study, we examine the dynamic properties of Weyl points in multi-terminal Josephson junctions.
The junctions are modulated by external parameters, such as electric gate voltage, magnetic flux, bias voltage.
The Weyl points are manipulated and draw trajectories accompanied by pair creation and annihilation.
The trajectories form both closed loops and open lines.
We classify these trajectories using the Chern number and the phase diagram.
\end{abstract}
\maketitle

\section{INTRODUCTION}
\label{sec:intro}

Topological property in condensed matter is one of the most attractive subjects in recent physics
since it provides robust (protected) electronic states~\cite{Hatsugai93,Murakami07}.
Such topological physics and protection are extended to photonics~\cite{Haldane08,LLu14},
electronic circuits~\cite{CHLee18,Yatsugi22,HYang24},
nonequilibrium phenomenon~\cite{Rudner20,Sone24}, etc.
Moreover, the application of topological protection is proposed, e.g.
for switching device~\cite{Gilbert21,HWang22}, quantum computing~\cite{Nayak08,Mong14}, etc.

Weyl semimetal is one of the topological materials, which has singular points of band touching in three-dimensional (3D) bulk band structure~\cite{XWan11,Burkov11,Singh12}.
The singular point possesses positive or negative topological charge
(geometrical monopole in the momentum space) generating the Berry curvature~\cite{Murakami07}.
This geometrical charge guarantees the topological protection of Weyl points (WPs).
The presence of WPs provides peculiar surface states, called Fermi arc~\cite{XWan11,SYXu15,SJia16},
according to the bulk-edge correspondence~\cite{Hatsugai93}.
The WPs contributes to a chiral anomaly under electric and magnetic fields~\cite{SJia16,XYuan20,NPOng21},
which gives rise to attractive electrical properties, such as negative magnetoresistance~\cite{Burkov15,XHuang15},
anomalous Hall~\cite{Zyuzin12}, and Nernst effects~\cite{CZeng22}, etc.
Namely, the WPs can raise bulk properties.

We investigate multi-terminal Josephson junction (MTJJ).
The MTJJ is an attractive system because it represents an extension of the conventional Josephson effect,
owing to non-local superconducting phases.
In recent experiments, significant phenomena were reported, e.g.,
nonlocal correlation of supercurrent~\cite{Cohen18,Draelos19,Pankratova20,Graziano20,Matsuo22NL,Graziano22},
entanglement of two Cooper pairs~\cite{KFHuang22},
multiplet supercurrent~\cite{Arnault22,FZhang23,Arnault25},
nonreciprocal supercurrent~\cite{Chiles23},
fractional Shapiro step~\cite{Arnault21,Matsuo25SS},
anomalous Josephson effect~\cite{Matsuo23AJE,Prosko24},
Josephson diode effect~\cite{Matsuo23JDE,Gupta23,Coraiola24JDE},
Andreev molecule~\cite{Matsuo23AM,Coraiola23AM,Haxell23AM},
zero energy states~\cite{Matsuo23AM,Strambini16,Coraiola24ZE},
crossed Andreev reflection~\cite{Bordin23}, etc.

In addition to such physics owing to the nonlocality, MTJJ can emerge the WPs in the 3D space by the superconducting phases~\cite{Yokoyama15,Riwar16,Eriksson17,Meyer17,HYXie17,HYXie18,Erdmanis18,HYXie19,Houzet19,Klees20,Klees21,Weisbrich21,YChen21A,YChen21B,Boogers22,Repin22,HYXie22,Gavensky23,Septembre23,Teshler23,Mukhopadhyay23,Matute-Canadas24,Frank24,Zalom24,Ohnmacht25,Ram25}.
The emerged Weyl physics in the MTJJ follows the protection by topological charge.
Although a surface state owing to the bulk-edge correspondence cannot be considered in the MTJJ,
several schemes to detect the Berry curvature in the MTJJ were proposed~\cite{Riwar16,Eriksson17,Klees20}.
Finite Berry curvature without the WPs was also considered~\cite{Deb18,Gavensky18,Repin19}.
The Weyl physics in Josephson junctions can be extended.
Several studies considered the WP emergence in the circuits with multiple Josephson junctions\cite{Stenger19,Fatemi21}.
The WPs emerge at zero energy.
It might be related to the Majorana fermion,
which is also created at zero energy at the end of a superconducting region in a 1D nanowire~\cite{Mourik12,Rokhinson12,Das12,Deng12}.
For the MTJJ, the presence of Majorana fermions was also studied~\cite{Zazunov17,Gavensky19,Sakurai20,TZhou20,Meyer21,Kenawy24}.
The presence of zero energy states is key ingredient to extend the Weyl physics in the MTJJs.
Then, the density of states at zero energy was investigated in several studies~\cite{vanHeck14,Padurariu15,Vischi17,Yokoyama17,Wisne24}.
SNSNS junction is broadly one of the MTJJ.
In the normal regions, topologically protected zero energy states can be formed~\cite{Lesser21,Lesser22},
which is slightly connected to the Majorana zero modes.
This zero energy state in the SNSNS junctions is related to the Andreev molecule state~\cite{Kornich19,Kornich20}.
Moreover, the MTJJ has rich attractive physics even without Weyl physics, e.g.,
Cooper quartet~\cite{Freyn11,Jacquet20,Melin20,Melin21,Melin22,Melo22,Melin23,Melin24,Cayao24,Ohnmacht24},
nonquiliblium phenomena~\cite{Melin16,Nowak19,Melin19},
fractional transconductance~\cite{Peyruchat21,Weisbrich23},
chiral transmission~\cite{Day25},
etc~\cite{Alidoust12,Mai13,Moor16,ZQi18}.

In this study, we examine a manipulation of the WPs in the 3D space of the superconducting phase $\{ \varphi_j \}$ in MTJJs.
For four-terminal junction, we consider mesoscopic systems for the normal region with quantum point contacts (QPCs) embedded between
the normal and superconducting regions.
The system is tuned electrically by the gate voltage on the QPCs.
By quenching the electron transport through the QPC, we can change the dimension of $\{ \varphi_j \}$-space continuously.
It promises a topological phase transition from topological state with four WPs to trivial state with no WPs~\cite{NielsenNinomiya}.
The manipulation of WPs is examined also by the superconducting phase for five-terminal junction.
The phase can be tuned by an external magnetic flux or a finite voltage, namely AC Josephson effect.
They break the time-reversal symmetry (TRS).
Hence the topological phase transition shows the different features from those by the electrical QPC tuning.
We focus on the trajectory of WPs~\cite{Yokoyama15,Septembre23,Frank24}.
Accompanying with the pair annihilation and creation of WPs, the WP trajectories form closed loops or open lines in
the tuning of QPC gate voltage and additional superconducting phase.
It enables us the classification of WP trajectories.
To support the classification, we examine the number of WPs as the phase diagram and the Chern number.
From the phase diagram, we find a different topological state with only two WPs under the TRS breaking.
The Chern number might be detectable as a transconductance~\cite{Riwar16}.
With the trajectories, we can consider novel topological classification of the dynamic properties.

The remainder of this article is structured as follows.
In Section II, we describe the model and formulation based on the scattering matrix~\cite{Yokoyama17}.
Section III is devoted to four-terminal junctions, where we discuss the closed and open trajectories under the TRS.
In Section IV, we consider five-terminal junctions.
Here, we find rich possibility of the topological states in the MTJJ.
Finally, we present discussion and conclusions in Section V.

\section{Model and Formulation}

In this section, we explain a model for the MTJJ with the QPCs based on the scattering matrix.
The normal region is described by a random matrix.
Following the Beenakker formulation~\cite{Beenakker91}, we evaluate the Andreev levels and discuss the WPs emerged in random samples.

\subsection{System}
\begin{figure}[t]
\includegraphics[width=80mm]{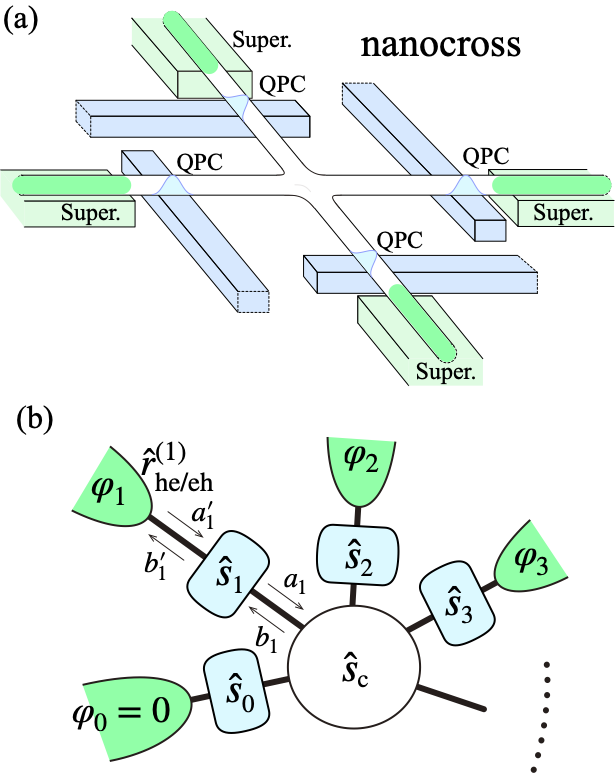}
\caption{Schematics of multi-terminal Josephson junction.
(a) An example of multi-terminal Josephson junction using semiconductor nanowires with the QPC structures.
(b) A model of multi-terminal Josephson junction depicted the scattering matrix.}
\label{fig:model}
\end{figure}

Figure \ref{fig:model}(a) depicts an example of nanostructured four-terminal junction using semiconductor nanocross.
Such multi-terminal structures could be fabricated by using
crossed nanowire~\cite{Plissard13}, quantum well~\cite{Pankratova20}, graphene~\cite{Draelos19,Arnault21}, etc.
In this schematic, four superconductors induce superconducting regions by the proximity effect.
We consider metallic gate electrodes beside the four superconductors.
The gate electrodes form potential barriers in the nanocross, which work as the QPCs,
and tune the transmission between the normal and superconducting regions.
For simplicity, we assume a single conduction channel through each QPC potential.
The Andreev reflection occurs at the boundaries between the normal and superconducting regions.

Let us describe the system in terms of the scattering matrix.
Figure \ref{fig:model}(b) schematically represents the scattering matrix description for the MTJJ.
In the central part of the normal region, electrons from each QPC with the coefficients $(a_{0,{\rm e}}, \cdots ,a_{N-1,{\rm e}})$ of
conduction channels are scattered to $(b_{0,{\rm e}}, \cdots ,b_{N-1,{\rm e}})$,
\begin{equation}
\vec{b}_{\rm e}
=
\hat{s}_{{\rm c,e}}
\vec{a}_{\rm e}.
\label{eq:Sce}
\end{equation}
Here, we introduce the vectors $\vec{a}_{\rm e} = (a_{0,{\rm e}}, \cdots , a_{N-1,{\rm e}})^{\rm t}$ and
$\vec{b}_{\rm e} = (b_{0,{\rm e}}, \cdots , b_{N-1,{\rm e}})^{\rm t}$.
For the QPC structure adjusting to the $j$-th superconductor, the matrix $\hat{s}_{j,{\rm e}}$ describes the scattering of
injected electrons with $(b_{j,{\rm e}} , a_{j,{\rm e}}^\prime)$ to ejected ones with $(a_{j,{\rm e}} , b_{j,{\rm e}}^\prime)$,
\begin{equation}
\left( \begin{matrix}
a_{j,{\rm e}} \\ b_{j,{\rm e}}^\prime
\end{matrix} \right)
=
\hat{s}_{j,{\rm e}}
\left( \begin{matrix}
b_{j,{\rm e}} \\ a_{j,{\rm e}}^\prime
\end{matrix} \right)
\label{eq:Sje}
\end{equation}
Here, $j=0 ,\cdots, N-1$ with $N$ being a number of superconducting terminals.
For holes with $a_{j,{\rm h}}, a_{j,{\rm h}}^\prime, b_{j,{\rm h}}, b_{j,{\rm h}}^\prime$,
the scattering matrices $\hat{s}_{{\rm c,h}}$ and $\hat{s}_{j,{\rm h}}$ are defined in the same manner as
Eqs.\ (\ref{eq:Sce}) and (\ref{eq:Sje}), respectively.
The Andreev reflection is also described in terms of the scattering matrix,
\begin{equation}
\vec{a}_{\rm h (e)}^\prime
=
\hat{r}_{\rm he (eh)}
\vec{b}_{\rm e (h)}^\prime
\label{eq:AR}
\end{equation}
with $\vec{a}_{\rm e}^\prime = (a_{0,{\rm e}}^\prime, \cdots , a_{N-1,{\rm e}}^\prime)^{\rm t}$ and
$\vec{b}_{\rm e}^\prime = (b_{0,{\rm e}}^\prime, \cdots , b_{N-1,{\rm e}}^\prime)^{\rm t}$.
The vectors for holes are defined in the same manner.
Here, the scattering matrix describing the Andreev reflection is a diagonal matrix;
\begin{equation}
\hat{r}_{\rm he(eh)} = {\rm diag} (r_{0,{\rm he(eh)}}, \cdots , r_{N-1,{\rm he(eh)}}).
\end{equation}
The elements include the superconducting phases $\varphi_j$ as $r_{\rm he}^{(j)} = e^{-i\varphi_j} e^{-i \arccos (E/\Delta_0)}$.
Here, $\Delta_0$ is a superconducting pair potential in the superconducting regions.
We set $\varphi_0 = 0$ without loss of generality.

\subsection{Combining of scattering matrix}
The size of the electron (hole) scattering matrix $\hat{s}_{\rm c,e (h)}$ for the central region is $N \times N$
when each QPC connects only a single conduction channel to the superconducting region.
We do not indicate the spin degrees of freedom explicitly because neither the Zeeman effect nor the spin-orbit interaction is considered.
The electron transport through the QPC is described by $\hat{s}_{j,{\rm e}}$.
It is characterized by the transmission probability~\cite{NazarovBlanter} as
\begin{equation}
\hat{s}_{j,{\rm e}} =
\left( \begin{matrix}
\sqrt{1 - T_j} e^{i \theta_j} & \sqrt{T_j}      e^{i \eta_j} \\
\sqrt{T_j}      e^{i \eta_j}    & -\sqrt{1 - T_j} e^{i (2\eta_j - \theta_j)}
\end{matrix} \right).
\end{equation}
The parameters $\eta_j$ and $\theta_j$ describe additional phases through the QPCs.
However, they are not essential since their effect can be incorporated into the randomness of $\hat{s}_{\rm c, e}$.

We construct the scattering matrix for the normal region by combining $\hat{s}_{\rm c,e}$ and $\hat{s}_{j,{\rm e}}$~\cite{Yokoyama17},
which satisfies a following form
\begin{equation}
\vec{b}_{\rm e}^\prime = \hat{s}_{\rm sys ,e} \vec{a}_{\rm e}^\prime.
\end{equation}
For the vectors of ``in-coming'' and ``out-going'' electron's coefficients, we can establish
\begin{equation}
\left( \begin{matrix}
\vec{b}_{\rm e}^\prime \\
\vec{a}_{\rm e} \\
\vec{b}_{\rm e}
\end{matrix} \right)
=
\left( \begin{matrix}
\hat{\Sigma}_{11} & 0                           & \hat{\Sigma}_{13} \\
\hat{\Sigma}_{21} & 0                           & \hat{\Sigma}_{23} \\
0                 & \hat{\Sigma}_{32}           & 0
\end{matrix} \right)
\left( \begin{matrix}
\vec{a}_{\rm e}^\prime \\
\vec{a}_{\rm e} \\
\vec{b}_{\rm e}
\end{matrix} \right).
\label{eq:allmatrix}
\end{equation}
The block matrices $\hat{\Sigma}_{kl}$ with $k,l = 1,2,3$ are given by the elements of scattering matrices, for instance,
$\hat{\Sigma}_{32} = \hat{s}_{\rm c}$.
By solving Eq.\ (\ref{eq:allmatrix}), we obtain
\begin{equation}
\hat{s}_{\rm sys ,e} = \hat{\Sigma}_{11} + \hat{\Sigma}_{13} \frac{1}{1 - \hat{\Sigma}_{32} \hat{\Sigma}_{23}} \hat{\Sigma}_{32} \hat{\Sigma}_{21}.
\end{equation}
By following this procedure~\cite{Yokoyama17}, we introduce the transmission probability $T_j$ to the normal region in the MTJJ,
and use it as a control parameter.
The scattering matrix for holes is given by the complex conjugate of that for electron because of the TRS,
\begin{equation}
\hat{s}_{\rm sys ,h} = \hat{s}_{\rm sys ,e}^\ast.
\end{equation}

By applying the matrices to the Beenakker formulation~\cite{Beenakker91},
\begin{equation}
\det \left( 1 - \hat{r}_{\rm eh} \hat{s}_{\rm sys ,e}^* \hat{r}_{\rm he} \hat{s}_{\rm sys ,e} \right) = 0,
\label{eq:Beenakker}
\end{equation}
and solving it, we obtain the energy of ABSs, $E_n (\vec{\varphi})$.

\subsection{Chern number}
Let us consider the eigenvectors of the ABS for short junction limit.
Assuming that all terminals have an equivalent superconducting gaps $\Delta_0$,
Eq.\ (\ref{eq:Beenakker}) is modified as
\begin{equation}
\det \left( e^{i 2\chi (E)} - \hat{S} (\vec{\varphi}) \right) =0
\label{eq:Beenakker2}
\end{equation}
with $\chi (E) = \arccos (E/\Delta_0)$ and
\begin{eqnarray}
\hat{S} (\vec{\varphi}) &=& \hat{s}^* (\vec{\varphi}) \hat{s} (\vec{\varphi}),
\label{eq:Smatrix} \\
\hat{s} (\vec{\varphi}) &\equiv & e^{-i \hat{\varphi}/2} \hat{s}_{\rm sys, e} e^{+i \hat{\varphi}/2}.
\label{eq:smalls}
\end{eqnarray}
Here, $\hat{\varphi}$ is a diagonal matrix by $\varphi_j$.
Based on Eq.\ (\ref{eq:Beenakker2}), the eigenstate for $E_n (\vec{\varphi})$ satisfies
\begin{equation}
e^{i 2\chi (E_n)} |\psi_n (\vec{\varphi}) \rangle  = \hat{S} (\vec{\varphi}) |\psi_n (\vec{\varphi}) \rangle 
\end{equation}
After short algebra, we find the effective Hamiltonian as
\begin{equation}
\hat{H} (\vec{\varphi}) \equiv \frac{1}{2} \left( \hat{S} (\vec{\varphi}) + \hat{S}^\dagger (\vec{\varphi}) + 2 \right)^{\frac{1}{2}}
\label{eq:effectiveHamiltonian}
\end{equation}
and the Schr\"{o}dinger equation
\begin{equation}
\hat{H} (\vec{\varphi}) |\psi_n (\vec{\varphi}) \rangle = E_n (\vec{\varphi}) |\psi_n (\vec{\varphi}) \rangle.
\label{eq:Schrodinger}
\end{equation}

This ``wave function'' $|\psi_n (\vec{\varphi}) \rangle$ provides the Berry curvature field,
\begin{equation}
\vec{B}_n (\vec{\varphi})
= i \left\langle \frac{\partial \psi_n}{\partial \vec{\varphi} } \right| \times \left| \frac{\partial \psi_n}{\partial \vec{\varphi} } \right\rangle.
\label{eq:Berry}
\end{equation}
The Berry curvature can be regarded as a field generated by the topological charges of the WPs.
The Chern number $Ch_{n}$ defined by the surface integral of Berry curvature~\cite{Berry84,Fukui05},
\begin{equation}
Ch_{n,3} (\varphi_3) = \frac{1}{2\pi} \int_{-\pi}^{\pi} d\varphi_1 \int_{-\pi}^{\pi} d\varphi_2 B_{n,3} (\vec{\varphi}),
\label{eq:Chern}
\end{equation}
becomes a discrete value and provides an index of the topological state.
In the following, we omit the subscript $n$ because we consider only the lowest energy band.

\section{Four-terminal junction}
First, we focus on a four-terminal case.
A tuning parameter is the transmission probability $T_{j=3}$ of the QPC for the fourth terminal. 
A decrease in $T_3$ from unity to zero shows the trajectories of the WPs and the topological phase transitions
accompanied by the pair annihilation and creation of the WPs.
An analysis of many samples leads to a classification of the trajectories.

\subsection{Pair of WPs}

\begin{figure}[t]
\includegraphics[width=80mm]{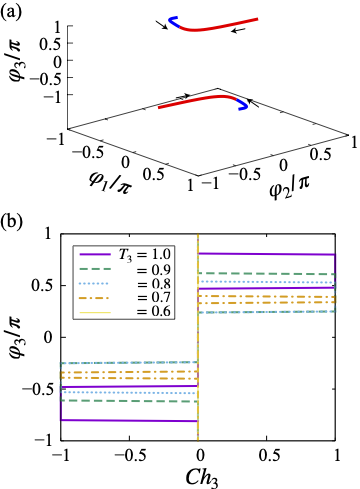}
\caption{The trajectory of WPs (a) and the Chern number (b) in one sample with the WPs at $T_3 = 1$.
With the decrease of $T_3$, only pair annihilation occurs at
$(\varphi_1,\varphi_2,\varphi_3) \approx \pm (0.762,-0.287,-0.366)\pi$ when $T_3 \approx 0.69$.}
\label{fig:pairannihilation}
\end{figure}

Figure \ref{fig:pairannihilation}(a) demonstrates a typical case with four WPs
at $\vec{\varphi} = \vec{\varphi}_{\rm W1/W2}^{(+)}$ for positive topological charge (red) and
at $\vec{\varphi}_{\rm W1/W2}^{(-)}$ for negative (blue line).
The four WPs are located at the TRS points ($\vec{\varphi}_{\rm W2}^{(\pm)} = -\vec{\varphi}_{\rm W1}^{(\pm)}$).
The positive or negative topological charge cause the Berry curvature
and provides discrete Chern number (Fig.\ \ref{fig:pairannihilation}(b)).
From the Chern number, we infer the sign of topological charge at the WPs~\cite{com1}.

Let us tune $T_3$ from 1 to 0.
At $T_3 = 0$, the system is effectively three terminals.
Hence, the $T_3$ tuning results in the annihilation of WPs as shown in Fig.\ \ref{fig:pairannihilation}(a).
Owing to the TRS holding under the $T_3$ tuning, the trajectories of WPs are symmetric with respect to the origin.
The Chern number is positive (negative) in $\varphi_3 = \varphi_{W1,3}^{(+)}$ to $\varphi_{W1,3}^{(-)}$
($\varphi_{W2,3}^{(-)}$ to $\varphi_{W2,3}^{(+)}$) in Fig.\ \ref{fig:pairannihilation}(b).
In this sample, at $T_3 \approx 0.69$, the two pair annihilations of WPs happen simultaneously.
Then, the topological phase transition only from four-WP phase to trivial phase is found.

In a case of no WPs at $T_3 = 1$, we can find the pair creation of WPs in the decrease of $T_3$, and they should annihilate.
Therefore, the WPs show continuous trajectories by $T_3$.
The trajectory strongly depends on the sample.
In the following, we examine a classification of the trajectories.

\subsection{Closed and open trajectories of WPs}

\begin{figure}[t]
\includegraphics[width=80mm]{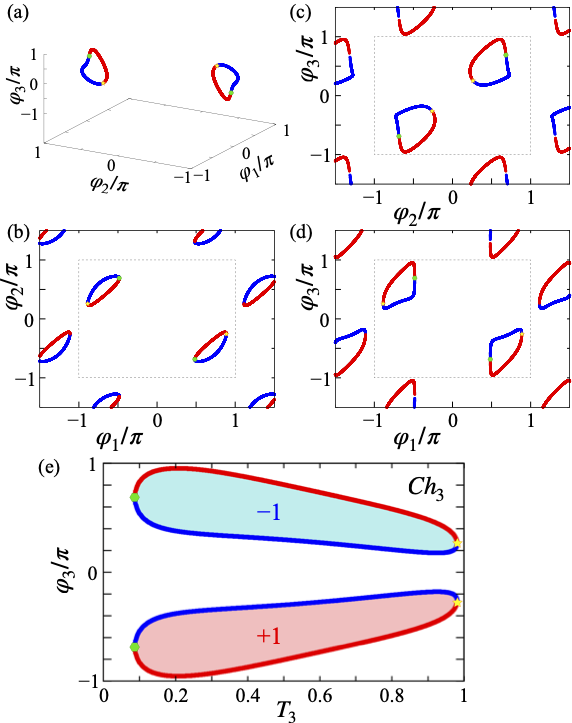}
\caption{Numerical results of a typical sample for four-terminal Josephson junctions.
The pair creation and annihilation of the WPs occur in the decrease of $T_3$.
The two WP pairs draw separated two closed trajectories without pair exchange, as shown in the 3D space (a),
in the projected 2D $\varphi_1$-$\varphi_2$ (b), $\varphi_2$-$\varphi_3$ (c), and $\varphi_1$-$\varphi_3$ planes (d).
Red and blue lines indicate the WP trajectories with positive and negative topological charge, respectively.
Yellow star and green hexagon marks indicate the positions of the pair creation and annihilation, respectively.
(e) Chern number is $Ch_3 = +1$ (light red) and $-1$ (light blue) between the projected positions of WPs in $\varphi_3$.}
\label{fig:4term1}
\end{figure}

\begin{figure}[t]
\includegraphics[width=80mm]{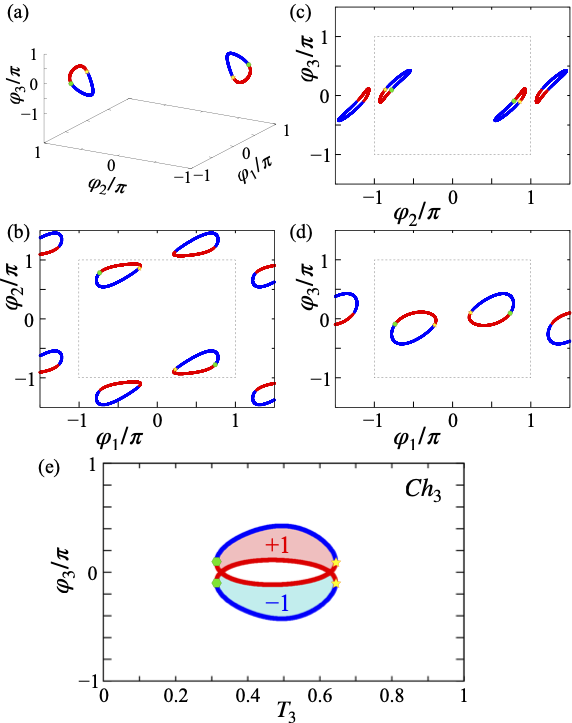}
\caption{Numerical results of another sample for closed two trajectories without pair exchange.
The plots in (a)-(e) are the same manner as those in Fig.\ \ref{fig:4term1}.
The regions of $Ch_3 = \pm 1$ are overlapped with each other, where $Ch_3 = 0$ by the cancellation.}
\label{fig:4term2}
\end{figure}

We examine many samples to investigate a classification of the WP trajectories.
Figures \ref{fig:4term1} and \ref{fig:4term2} demonstrate typical cases of the WP trajectory and the Chern number by a tuning of $T_3$.
Red and blue lines indicate (projected) positions of the WPs with positive and negative topological charge, respectively.
We pick up the samples being topologically trivial at $T_3 =1$ and with the pair creation and annihilation in the decrease of $T_3$.
Yellow star and green hexagon marks indicate the pair creation and annihilation, respectively.
The two pair creations and annihilations occur simultaneously.
The WP trajectories draw two separated closed loops, as shown in projected views in
Figs.\ \ref{fig:4term1}(b)-(d) and \ref{fig:4term2}(b)-(d).
Note that the trajectories can pass the boundary of Brillouin zone even for the small closed loop.
In the presented samples, however, such passing trajectory is not obtained.
Figures \ref{fig:4term1}(e) and \ref{fig:4term2}(e) show the Chern number $Ch_3$ in the $\varphi_3$-direction.
$Ch_3 = \pm 1$ is found in the interval of pair WPs, $\varphi_{W1,3}^{(+)}$ to $\varphi_{W1,3}^{(-)}$ for positive and
$\varphi_{W2,3}^{(-)}$ to $\varphi_{W2,3}^{(+)}$ for negative.
In Fig.\ \ref{fig:4term1}, the two trajectories are well separated, hence the intervals of $Ch_3 = \pm 1$ are also separated,
while the sample of Fig.\ \ref{fig:4term2} shows an overlap of positive and negative $Ch_3$ intervals
although the closed loop trajectories are still separated well in the 3D space.
By the cancellation, the overlap region indicates $Ch_3 = 0$.

\begin{figure}[t]
\includegraphics[width=80mm]{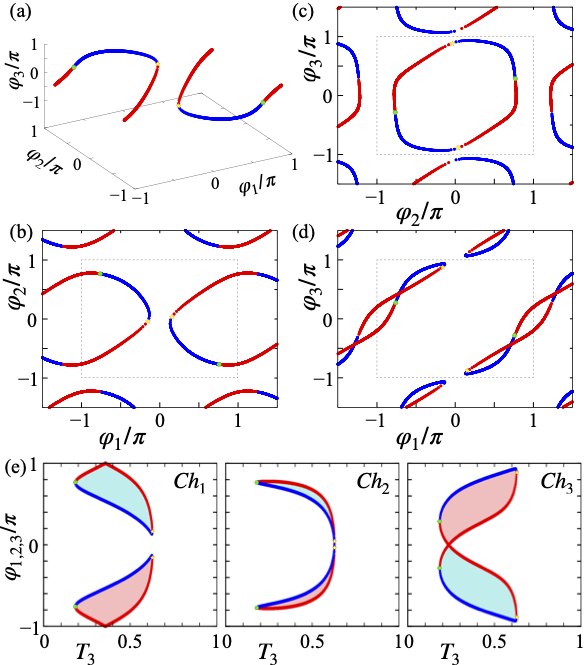}
\caption{Numerical results of a sample with pair exchange of WPs.
The plots in (a)-(e) are the same manner as those in Fig.\ \ref{fig:4term1}.
The trajectory draws one closed loop.}
\label{fig:4term3}
\end{figure}

The samples of Figs.\ \ref{fig:4term1} and \ref{fig:4term2} indicate pair creation and annihilation without pair exchange.
On the other hand, in Fig.\ \ref{fig:4term3}, we find the pair exchange, and the four WPs form one closed loop.
Here, let us call the two WP pairs created at $\vec{\varphi} = \vec{\varphi}_{\rm W1}^{(\pm)}$ and
$\vec{\varphi}_{\rm W2}^{(\pm)}$ (yellow stars) as Pair 1 and Pair 2.
With the decrease of $T_3$, the pair WPs are far from each other.
Then, the negative WP of Pair 1 (Pair 2) is closer to the positive WP of Pair 2 (Pair 1).
Note that the positive WPs go out to the neighboring Brillouin zone and their copies come from the opposite neighboring zone.
As a result, the positive WP of Pair 1 (Pair 2) annihilates with the copy of negative WP of Pair 2 (Pair 1) at green hexagon marks.

The feature of the pair exchange is found also in the Chern number.
For two separated loops, in Fig.\ \ref{fig:4term2}(e),
$Ch_3 = \pm 1$ regions cross twice (even times) with each other in $T_3$ at $\varphi_3 = 0$ or $\pm \pi$,
whereas, for single loop, the crossing of $Ch_j = \pm 1$ regions is once in $T_3$ decreasing, as shown in Fig.\ \ref{fig:4term3}(e).
Note that for $Ch_2$, the touching occurs at $\varphi_2 = 0$.

\begin{figure}[t]
\includegraphics[width=80mm]{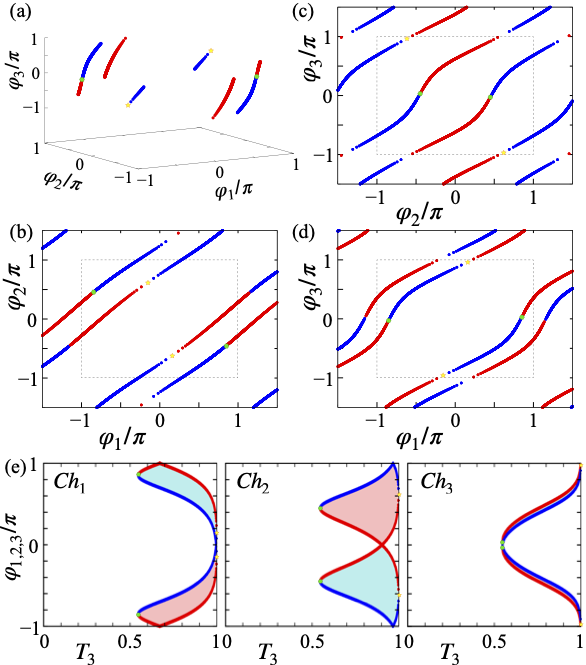}
\caption{Numerical results of a sample for open trajectory with no pair exchange
but the pair annihilation with a copy in the neighboring Brillouin zone.
The plots in (a)-(e) are the same manner as those in Fig.\ \ref{fig:4term1}.}
\label{fig:4term4}
\end{figure}

In addition, we can find another type of trajectory in Fig.\ \ref{fig:4term4}.
In this case, the created WPs are annihilated without pair exchange, but with the copy of itself at the neighboring Brillouin zone.
Therefore, the WPs demonstrated two open trajectories as shown in Figs.\ \ref{fig:4term4}(b)-(d).
In Fig.\ \ref{fig:4term4}(e), the crossing of $Ch_3 = \pm 1$ regions is found once at $\varphi_j = 0$ and once at $\pm \pi$.

\subsection{Statistic data}

Our approach is based on the random matrix.
Then, we demonstrate 100,000 samples to follow the kinds of WP trajectories accompanying with the pair creation and annihilation.
In the samples, we find above five cases: no WP, only pair annihilation (WPs exist at $T_4 = 1$, Fig.\ \ref{fig:pairannihilation}),
two closed trajectory (Figs.\ \ref{fig:4term1} and \ref{fig:4term2}), single closed loop by 4 WPs (Fig.\ \ref{fig:4term3}),
open trajectory (Fig.\ \ref{fig:4term4}).
In addition to them, there are $210$ undeterminable samples owing to insufficient position data for the WP trajectory
(Almost of them might be the case of two closed trajectory).
In Table \ref{tab:4term}, we summarize the statistic data of demonstration.
In previous study~\cite{Yokoyama15}, the emergence of WPs is evaluated only at $T_j =1$.
However, in the tuning of $T_3$, we find the pair creation of WPs, hence the probability of WP emergence is higher.

\begin{table}[b]
\begin{tabular}{c|c|c|c|c||c}
no WP  & annihilation & two closed  & single closed & open  & sum \\
\hline \hline
87,831 & 4,079        & 4,211 + 210 & 356           & 3,313 & 100,000
\end{tabular}
\caption{Statistic data of the trajectories for four-terminal junctions.}
\label{tab:4term}
\end{table}

\section{Five-terminal systems}

Next, we consider an extension of the systems to five-terminal case, where
the additional superconducting phase $\varphi_4$ and the transmission probability $T_4$ of the fifth-terminal can be
magnetically and electrically well-controllable parameters, respectively.
The two parameters are tuned periodically.
Owing to the AC Josephson effect under a finite bias voltage, $\varphi_4$ could be evolved electrically in time.

For a five-terminal junction, we can consider a 4D space spanned by
$(\varphi_1,\varphi_2,\varphi_3,\varphi_4)$ for the energy of ABSs~\cite{Riwar16}, where the Weyl singularity becomes a 1D line.
Such higher-dimensional picture would provide a platform to realize novel emergent physics.
However, in this paper, we treat $T_4$ and $\varphi_4$ as control parameters for the WPs in the 3D space by $(\varphi_1,\varphi_2,\varphi_3)$.
We discuss the phase diagram for the topological states in a $\varphi_4$-$T_4$ plane, which exhibits the topological characteristics.
In addition, the Chern number $Ch_3$ in $\varphi_4$ cross section at fixed $T_4$ shows several features.
From these analyses, we discuss a classification of the WP trajectory in five-terminal junction.

\subsection{Phase diagram by numuber of WPs}

Let us examine the phase diagram and the Chern number in Figs.\ \ref{fig:5term1} and \ref{fig:5term2},
and additionally in Figs.\ A1-A4 in Appendix.
The phase diagram is examined by the number of WPs in the $\varphi_4$-$T_4$ plane.
Here, $\varphi_4$ can be swept in time by the AC Josephson effect.
Note that $\varphi_4$ is swept from $-\pi$ to $\pi$ for readability to see the TRS relations.

\begin{figure}[t]
\includegraphics[width=80mm]{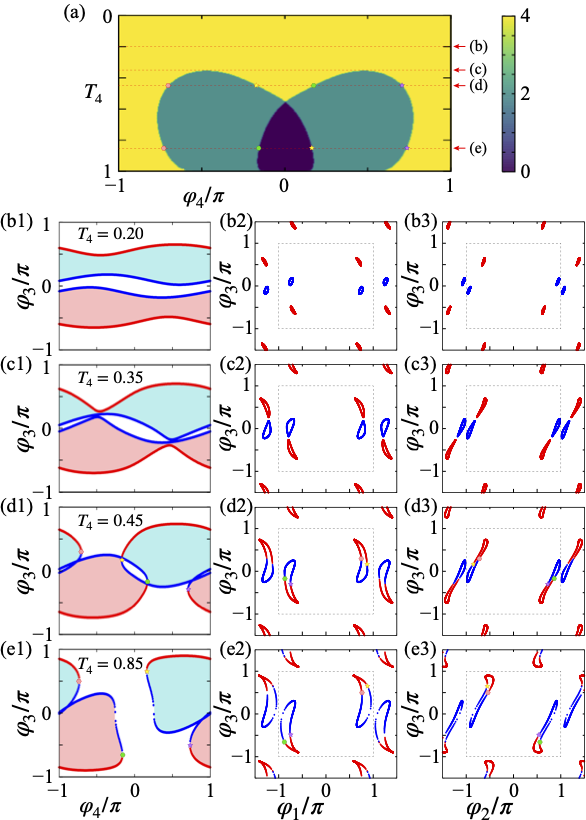}
\caption{Numerical results of a typical sample for five-terminal Josephson junction.
(a) Number of the WPs as the phase diagram in the $\varphi_4$-$T_4$ plane.
Four WPs are found at $T_4 = 0$ (four-terminal set-up).
Red broken lines are reference to indicate $T_4 = 0.20$ (b), $0.35$ (c), $0.45$ (d), $0.85$ (e).
(b) Cross-sectional plot at fixed $T_4 = 0.20$ for the Chern number $Ch_3$ (b1) and
the projected WP trajectory on $\varphi_1$-$\varphi_3$ (b2) and $\varphi_2$-$\varphi_3$ plane (b3).
(c)-(e) Results of $Ch_3$ and the WP trajectory at $T_4 = 0.35$, $0.45$, and $0.85$, respectively, with the same manner as those in (b).
In (a), (d), and (e), yellow and purple stars indicate the pair creation and pink and green hexagons are the pair annihilation
in the increase of $\varphi_4$.}
\label{fig:5term1}
\end{figure}

Figure \ref{fig:5term1}(a) exhibits the phase diagram.
In this sample, four WPs are obtained at small $T_4$, where $\varphi_4$ rarely affects the system.
Hence the WPs indicate individual small oscillation in the $\varphi_4$ sweep [Fig.\ \ref{fig:5term1}(b)].
The region of $Ch_3 = \pm 1$ shifts in $\varphi_3$ only slightly, and no overlap of $Ch_3 = \pm 1$ in Fig.\ \ref{fig:5term1}(b1).
The projected WP trajectories in Figs.\ \ref{fig:5term1}(b2) and (b3) show isolated oscillations of the four WPs.
Such isolated oscillation without pair creation and annihilation is not considered in the previous section for four-terminal case
because the pair annihilation must occur in the decrease of $T_3$ and the WPs must disappear at $T_3 = 0$.

With an increase of $T_4$, the trajectory of WPs becomes larger.
At $T_4 = 0.35$ [Fig.\ \ref{fig:5term1}(c)], two WPs with negative topological charge
intersects with each other in the $\varphi_3$ direction, and $Ch_3 = \pm 1$ regions overlap at $\varphi_3 \approx 0$.
At the overlap region, the Chern number becomes zero, like as that in Fig.\ \ref{fig:4term2}(e).
Let me note that for many samples with up to four WPs, we do not find $Ch_j = \pm 2$.
The trajectories of positive and negative charge are closer to each other.
However, the pair annihilation (and creation) is not found.

At $T_4 = 0.45$, we find the pair creation and annihilation by $\varphi_4$, which
are marked by star and hexagon, respectively, in Fig.\ \ref{fig:5term1}(d).
Owing to the TRS, they occur at sign flipped phases,
$(\varphi_1^{\rm (c)},\varphi_2^{\rm (c)},\varphi_3^{\rm (c)},\varphi_4^{\rm (c)})
= (-\varphi_1^{\rm (a)},-\varphi_2^{\rm (a)},-\varphi_3^{\rm (a)},-\varphi_4^{\rm (a)})$.
Hence the creation and annihilation do not occur at the same time in $\varphi_4$ except at $\varphi_4 = 0, \pm \pi$.
Here, we find a significant difference from the four-terminal junctions.
In Figs.\ \ref{fig:pairannihilation}-\ref{fig:4term4}, two pair annihilation (creation) occur at the same time in $T_3$,
whereas in Fig.\ \ref{fig:5term1}(d), the two pair annihilation marked by green and pink hexagons
(creation by yellow and purple stars) occur at different $\varphi_4$.
Hence, the number of WPs changes from 4 to 2 (from 2 to 4), and we find a topological phase with two WPs.
In this sample, the pair exchange of WPs, such as Fig.\ \ref{fig:4term3}) does not occur.

At larger $T_4$ in Fig.\ \ref{fig:5term1}(e), the position of pair creation and annihilation shifts and
we obtain the trivial phase (no WPs) around $\varphi_4 =0$.

\subsection{Multiple pair creation and annihilation}

\begin{figure}[t]
\includegraphics[width=90mm]{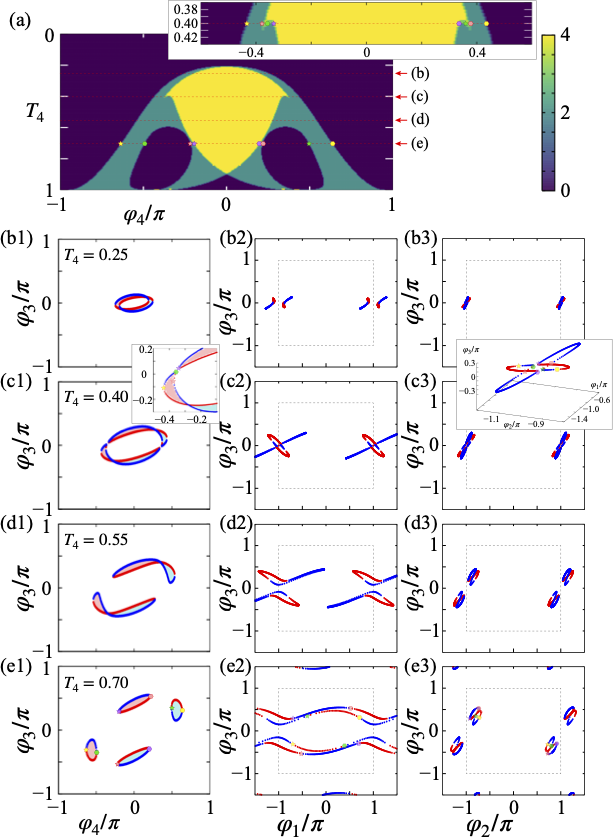}
\caption{Numerical results of another sample for five-terminal junction.
(a) Number of the WPs in the $\varphi_4$-$T_4$ plane. No WPs are found at $T_4 = 0$ (four-terminal set-up).
Red broken lines indicate $T_4 = 0.25$ (b), $0.40$ (c), $0.55$ (d), $0.70$ (e).
Inset is a zoom-up view for $T_4 \approx 0.40$.
(b)-(e) Cross-sectional plots for $Ch_3$ and the WP trajectory with the same manner as those in Fig.\ \ref{fig:5term1}.
In (a), (c), and (e), stars and hexagons indicate the pair creation and annihilation, respectively, in the increase of $\varphi_4$.
For (c), zoom-up views of (c1) and the 3D plot of trajectory are shown for visibility with star and hexagon marks.}
\label{fig:5term2}
\end{figure}

Let us consider another sample in Fig.\ \ref{fig:5term2}, where no WPs are found at small $T_4$.
At $T_4 = 0.25$, the phases of two and four WPs are obtained in Fig.\ \ref{fig:5term2}(a).
It is also found in the plot of Chern number.
In Fig.\ \ref{fig:5term2}(b1), two pair creations occur subsequently at $\varphi_4 \approx -0.245\pi$ and $\approx -0.190\pi$.
Their annihilations occur at $\varphi_4 \approx 0.190\pi$ and $\approx 0.245\pi$, respectively.
The WP pair from $\varphi_4 \approx -0.245\pi$ to $0.190\pi$ ($-0.190\pi$ to $0.245\pi$) results in positive (negative) $Ch_3$. 
However, their trajectories are close to each other [Figs.\ \ref{fig:5term2}(b2) and (b3)],
hence finite $Ch_3$ is found in tiny regions in $\varphi_3$.

At $T_4 = 0.40$ in Fig.\ \ref{fig:5term2}(c), the WP trajectories of two pair are combined, and the pair exchange occurs.
Then, additional pair annihilation and creation are found [see an inset of Fig.\ \ref{fig:5term2}(c1)].
The projected plots in Figs.\ \ref{fig:5term2}(c2) and (c3) exhibit the combination of the WP trajectories.
Let us show its 3D profiles in the inset.

With an increase of $T_4$, the combined trajectory by four WPs are separated to two trajectories and form new two pairs of the WPs.
It is clearly found by a comparison between the trajectories at $T_4 = 0.25$ in Fig.\ \ref{fig:5term2}(b)
and at $T_4 = 0.55$ in Fig.\ \ref{fig:5term2}(d), where the projected trajectories are separated in
$\varphi_1$ and $\varphi_3$, respectively.

Interestingly, at $T_4 = 0.70$, another pair annihilation and creation occur additionally in Fig.\ \ref{fig:5term2}(e).
Here, the WPs emerge intermittently at $\varphi_4 \approx -0.630$ to $-0.495$, $-0.215$ to $0.215$, and $0.495$ to $0.630$.
The additional pair annihilation and creation do not cause the pair exchange,
but change the closed trajectory to the open one [see Fig.\ \ref{fig:5term2}(e2)].
Note that even with the multiple pair creation and annihilation, the number of WPs does not exceed four at fixed $\varphi_4$.
From Fig.\ \ref{fig:5term2}(d2) to (e2), each closed loop combines with itself, and forms two open lines.
Hence, the number of open line trajectories is four in contrast to two lines in Fig.\ \ref{fig:4term4}.

At $T_4 > 0.9$, four WPs state disappears (not shown in Fig.\ \ref{fig:5term2}).
However, the trajectory is not changed qualitatively from those at $T_4 = 0.70$.

\begin{table*}[t]
\begin{tabular}{c||c|c|c|c|c|c|c|c|c|c}
number of WPs        & \ no WP \  & \multicolumn{2}{c|}{four WPs}  &
  \multicolumn{6}{c|}{pair creation/annihilation} & \ multiple crea./anni. \ \\
\hline
topo./trivial state  &   0        & \multicolumn{2}{c|}{4}         &
  \multicolumn{2}{c|}{$0 \Leftrightarrow 2$} & \multicolumn{2}{c|}{\ \ \ \ \ $2 \Leftrightarrow 4$ \ \ \ \ \ } &
  \multicolumn{3}{c}{$0 \Leftrightarrow 2 \Leftrightarrow 4$} \\
\hline
trajectory           &   ---      & 4 loops     & 4 lines          &
  2 loops & 4 lines & \multicolumn{2}{c|}{2 loops} & 2 loops & 4 lines & complicated
\end{tabular}
\caption{Classification of the trajectories for five-terminal junctions.}
\label{tab:5term}
\end{table*}

\subsection{Classification}

The phase diagram in the ${\varphi_4}-{T_4}$ plane could be more complicated than
Figs.\ \ref{fig:5term1}(a) and \ref{fig:5term2}(a).
We examine 50,000 samples.
Pair exchange, as seen in Figs.\ \ref{fig:5term2}(c), is found in many samples.
In these cases, multiple pair creation and annihilation occur, making the phase diagram and the Chern number profile more complicated.
To see them, let us demonstrate several samples additionally in Appendix (Fig.\ A1-A6).
Figure S1-S3 demonstrate the samples of the pair exchange in closed loop, open line, and complicated trajectories, respectively.
Figure S4 demonstrates the sample with a large closed trajectory.
Owing to such complicated phase diagram and pair creation and annihilation in the five-terminal junctions, however,
it is difficult to define and count exactly the kinds of WP trajectories.
Hence, we just classify the WP trajectories.

When $T_4$ is fixed and $\varphi_4$ is swept, we find three cases:
no WP without creation, four WPs without annihilation, and WPs with pair creation and annihilation.
The third case is sub-classified for three cases:
($0 \Leftrightarrow 2$), ($2 \Leftrightarrow 4$), and ($0 \Leftrightarrow 2 \Leftrightarrow 4$) WPs.
The number of pair creation and annihilation is not fixed for each sub-classification.
In addition to the classification by the number of WPs, we could consider the trajectory like as Table \ref{tab:4term}.
From the phase diagram, however, it is not possible to declare the open or closed trajectory.
Even for the case of four WPs without the annihilation,
we find both the closed loop and open line trajectories in Figs.\ A5 and A6 in Appendix.
In these cases, four independent WPs form four loops and four lines, respectively.
For ($0 \Leftrightarrow 2$) case with the pair annihilation and creation,
we find 2 closed loops ($T_4 = 0.35$ in Fig.\ A4) and 4 open lines (high $T_4$ in Fig.\ \ref{fig:5term2}).
For ($2 \Leftrightarrow 4$), we find 2 closed loops trajectory ($T_4 = 0.45$ in Fig.\ \ref{fig:5term1}) but no open line.
In Fig.\ A6, the closed loop trajectory at $T_4 \le 0.74$ becomes the open line trajectory at $T_4 \ge 0.75$ by vanishing the two WPs state.
For the case of ($0 \Leftrightarrow 2 \Leftrightarrow 4$), the closed and open trajectories are found in
Figs.\ \ref{fig:5term1} and \ref{fig:5term2}, respectively.
In addition, the pair exchange can be obtained when the multiple pair creation and annihilation occur ($T_4 = 0.40$ in Fig.\ \ref{fig:5term2}).
The pair exchange tends to make the trajectory of WPs complicated, as shown in the inset.
In the five-terminal junctions, we do not find a single closed trajectory with the pair exchange in our demonstration.

We summarize those classification in Table \ref{tab:5term}.

\section{Discussion and Conclusions}
\label{sec:conclusions}
We have investigated the emergence of WPs and their trajectories by tuning the controllable parameters of Josephson junctions.
For four-terminal junctions, we examine the WP trajectory by the QPC gate voltage in one of the four terminals.
Then, we find four kinds of trajectories.
Especially, the closed and open trajectories demonstrate significant features.
For five-terminal junctions, we examine the trajectory of WPs by the controlled phase difference.
Although the trajectories are point symmetric with respect to the origin in the $(\varphi_1,\varphi_2,\varphi_3)$-space,
the position of WPs can be asymmetric at fixed $\varphi_4$ owing to the TRS breaking.
Then, we find a topological state with only two WPs, accompanied by a single pair creation or annihilation.

We have examined the classification of WP trajectory and phase diagram.
The closed and open trajectories indicate qualitative difference with each other,
such as the pair exchange, the pair annihilation with the WP originated from the neighboring Brillouin zone.
The trajectory of WPs and their pair creation and annihilation positions have to be point symmetric with respect to
$\varphi_{1,2,3} = 0$ owing to the TRS although it strongly depends on the sample.
Several studies have indicated the position of WPs in the parameter space clearly by considering well-defined system,
such as quantum dot~\cite{Klees20}, symmetric circuit~\cite{Fatemi21}, etc.
Our analysis based on the random matrix exhibits a rich variety of WP trajectories.
The WPs posses the topological charge, hence the dynamics of WPs along the trajectory means a topological charge flow.
Then, the classification of such various trajectories would extend the topological analysis.
It is beyond the scope of this study, and will be discussed in our next paper.

In an MTJJ, the Weyl physics can be tuned over a wide range of controllable parameters.
Then, we continuously obtain the phase transition between 4 WPs state and 2 WPs state in addition to the trivial state (no WPs).  
In 3D solid Weyl semimetals, such different Weyl phases require different symmetry classes to emerge 4 and 2 WPs~\cite{Murakami07,Zyuzin12symmetry}.
In addition to such advantages, the Weyl physics in MTJJ has a possibility to extend the symmetry class.
For example, type-II Weyl semimetal has tilted Weyl corn~\cite{Soluyanov15}.
In our set-up based on the $s$-wave superconducting pair potential for the terminals,
the electron-hole symmetry guarantees the mirror symmetry of Andreev bound states in energy, hence a type-II WP does not emerge.
An application of the anisotropic superconductor, such as $p$- and $d$-wave, might cause the type-II WP.
In addition to the structure modulation, material consideration can expand the topological physics emergence in the MTJJ.

\begin{acknowledgments}
K.T. is supported by JST, the establishment of university fellowships towards the creation of science technology innovation,
Grant Number JPMJFS2125
\end{acknowledgments}

\appendix
\renewcommand{\figurename}{Fig.\ A}
\setcounter{figure}{0}
\section{Additional results for five-terminal junctions}

In this Appendix, we demonstrate additional results for five-terminal Josephson junctions.

\begin{figure*}[th]
\includegraphics[width=100mm]{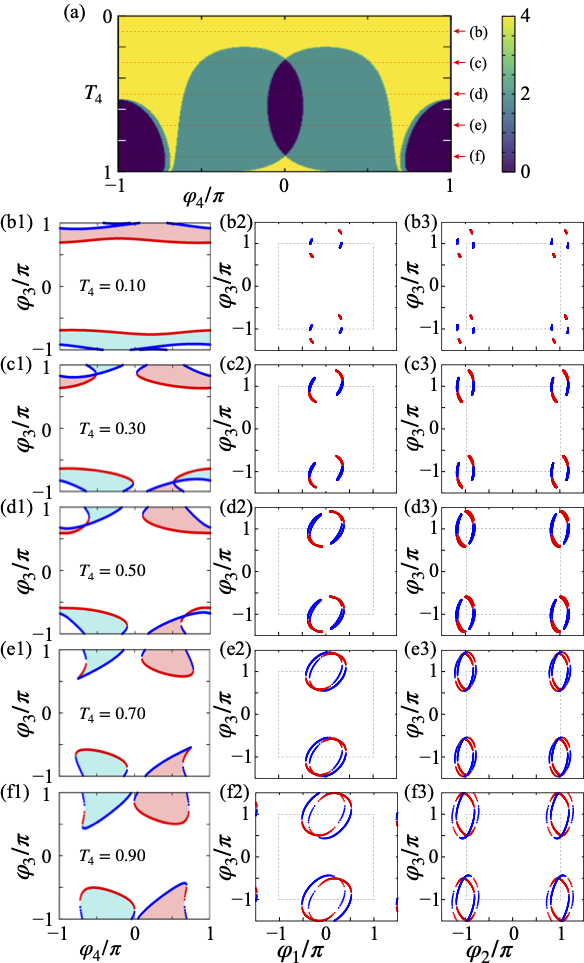}
\caption{Numerical results of a sample with pair exchange in closed loop trajectories.
(a) Number of the WPs as the phase diagram in the $\varphi_4$-$T_4$ plane.
Four WPs are found at $T_4 = 0$.
Red broken lines are reference to indicate $T_4 = 0.10$ (b), $0.30$ (c), $0.50$ (d), $0.70$ (e), $0.90$ (f).
(b) Cross-sectional plot at fixed $T_4 = 0.10$ for the Chern number $Ch_3$ (b1) and
the projected WP trajectory on $\varphi_1$-$\varphi_3$ (b2) and $\varphi_2$-$\varphi_3$ plane (b3).
(c)-(f) Results of $Ch_3$ and the WP trajectory at $T_4 = 0.30$, $0.50$, $0.70$, and $0.90$,
respectively, with the same manner as those in (b).}
\label{fig:A1}
\end{figure*}

Figure A\ref{fig:A1} exhibits the sample with the pair exchange in the closed loop trajectories and with four WPs at $T_4 = 0$.
The phase diagram, the Chern number plots, and the loop trajectories are qualitatively similar to those in Fig.\ 7 in the main text.
However, at $T_4 > 0.55$, multiple pair creation and annihilation is found, where the pair exchange of WPs occurs.

\begin{figure*}[th]
\includegraphics[width=100mm]{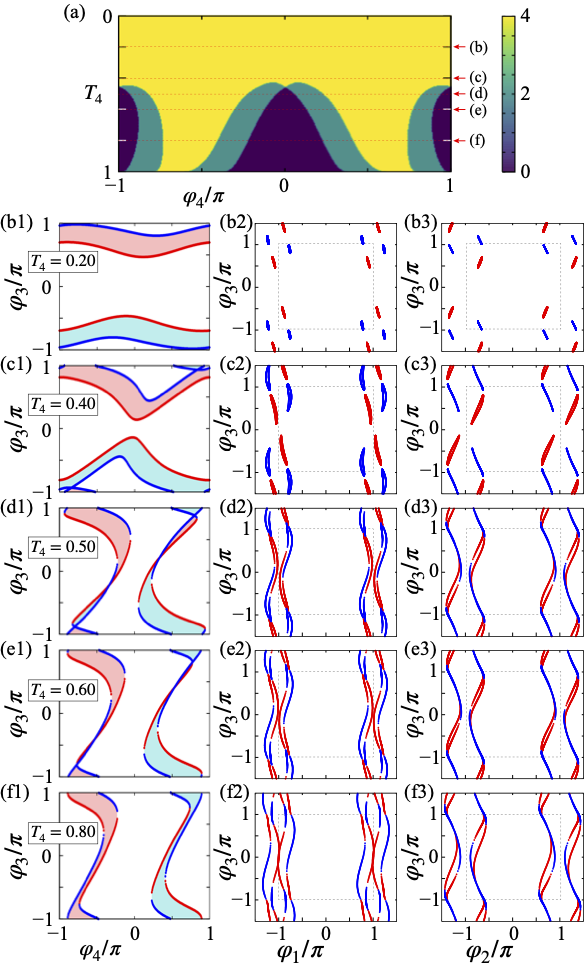}
\caption{Numerical results of a sample with pair exchange in open line trajectories.
Four WPs are found at $T_4 = 0$.
The parameters are the same as those in Fig.\ A\ref{fig:A1} except the transmission probability:
$T_4 = 0.20$ (b), $0.40$ (c), $0.50$ (d), $0.60$ (e), $0.80$ (f).}
\label{fig:A2}
\end{figure*}

Figure A\ref{fig:A2} is the sample with the pair exchange in open line trajectories and with four WPs at $T_4 = 0$.

\begin{figure*}[th]
\includegraphics[width=100mm]{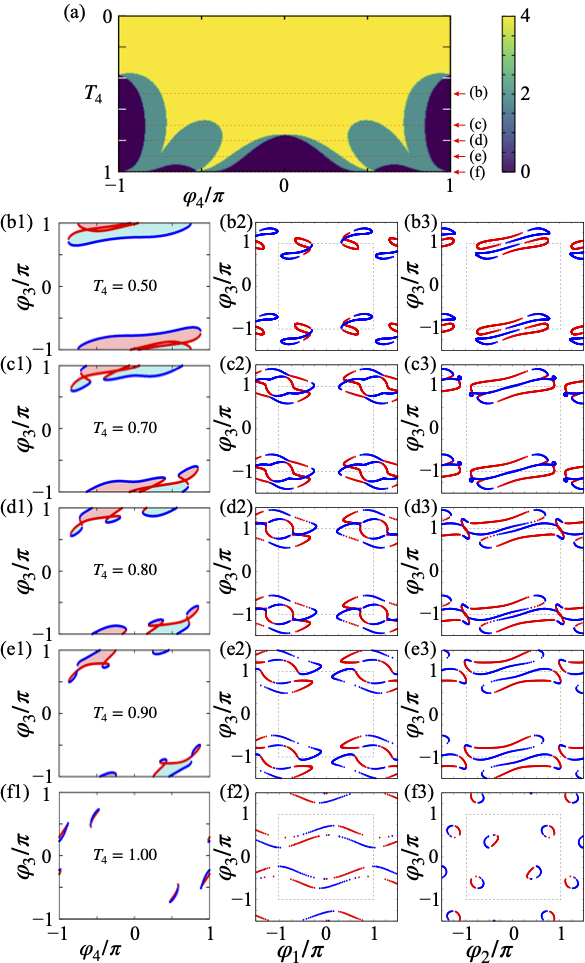}
\caption{Numerical results of a sample with pair exchange in complicated trajectories.
Four WPs are found at $T_4 = 0$.
The parameters are the same as those in Fig.\ A\ref{fig:A1} except the transmission probability:
$T_4 = 0.50$ (b), $0.70$ (c), $0.80$ (d), $0.90$ (e), $1.00$ (f).}
\label{fig:A3}
\end{figure*}

Figure A\ref{fig:A3} is the sample with the pair exchange in complicated trajectories and with four WPs at $T_4 = 0$.

\begin{figure*}[th]
\includegraphics[width=100mm]{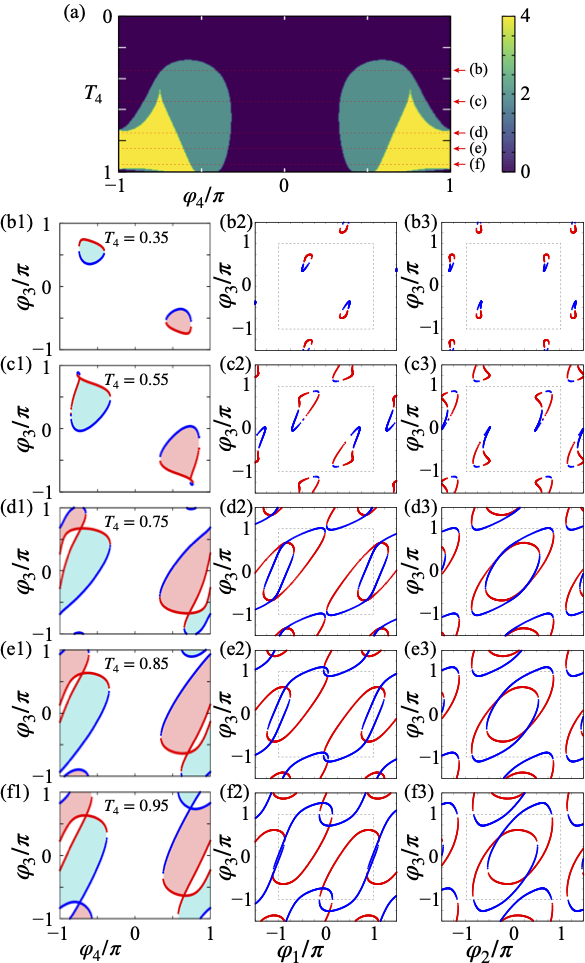}
\caption{Numerical results of a sample with large closed trajectories.
No WPs are found at $T_4 = 0$.
The parameters are the same as those in Fig.\ A\ref{fig:A1} except the transmission probability:
$T_4 = 0.35$ (b), $0.55$ (c), $0.75$ (d), $0.85$ (e), $0.95$ (f).}
\label{fig:A4}
\end{figure*}

Figure A\ref{fig:A4} is the sample with large closed trajectories and with no WP at $T_4 = 0$.
The large loop can surround the origin in the space by $\varphi_{1,2,3}$.

\begin{figure*}[th]
\includegraphics[width=100mm]{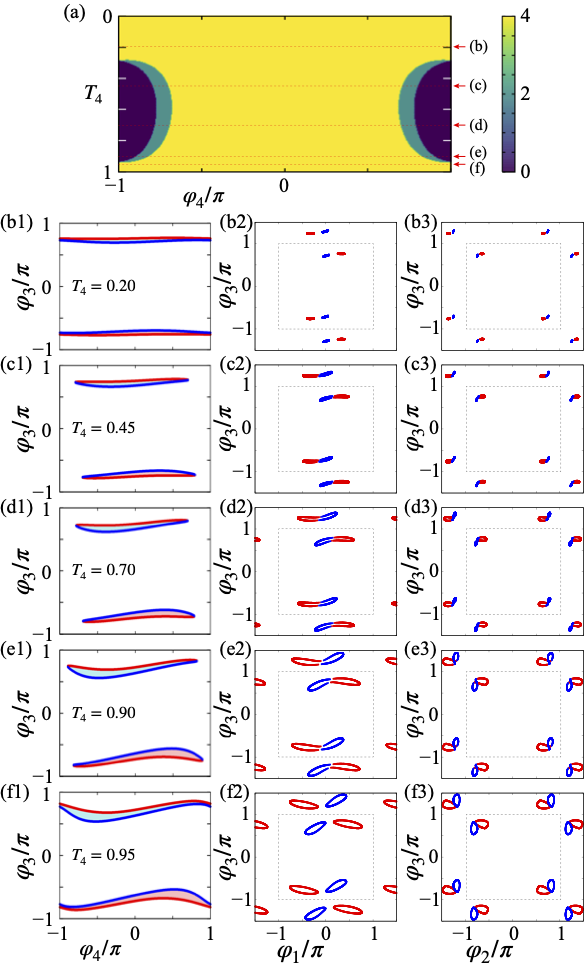}
\caption{Numerical results of a sample with closed trajectories without pair creation/annihilation.
Four WPs are found at $T_4 = 0$ and $1$.
The parameters are the same as those in Fig.\ A\ref{fig:A1} except the transmission probability:
$T_4 = 0.20$ (b), $0.45$ (c), $0.70$ (d), $0.90$ (e), $0.95$ (f).}
\label{fig:A5}
\end{figure*}

Figure A\ref{fig:A5} is the sample with independent four closed trajectories without pair creation/annihilation.

\begin{figure*}[th]
\includegraphics[width=100mm]{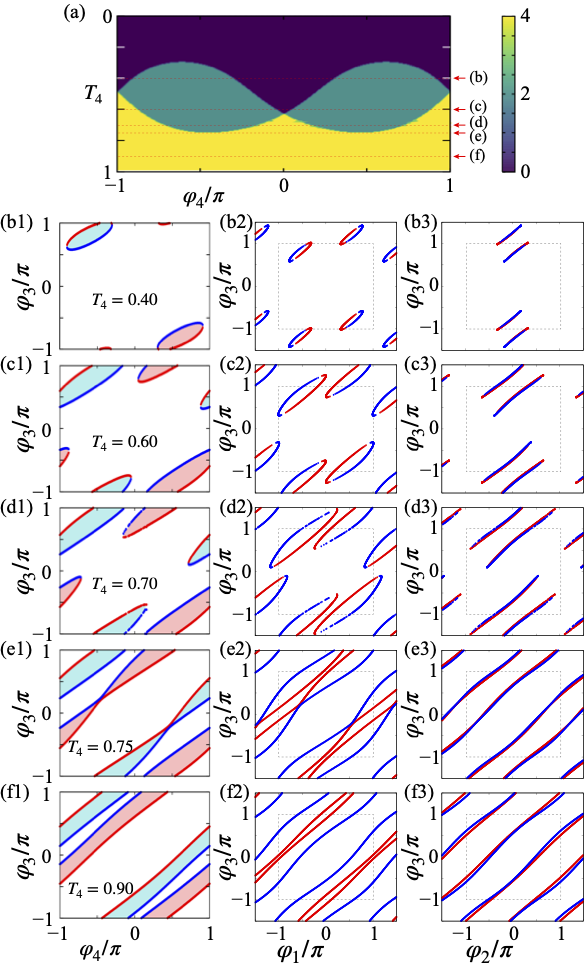}
\caption{Numerical results of a sample with open trajectories without pair creation/annihilation.
No WPs are found at $T_4 = 0$ and four WPs at $T_4 = 1$.
The parameters are the same as those in Fig.\ A\ref{fig:A1} except the transmission probability:
$T_4 = 0.40$ (b), $0.60$ (c), $0.70$ (d), $0.75$ (e), $0.90$ (f).}
\label{fig:A6}
\end{figure*}

Figure A\ref{fig:A6} is the sample with independent four open trajectories without pair creation/annihilation.

\begin{thebibliography}{99}

\bibitem{Hatsugai93} Y.\ Hatsugai, Phys.\ Rev.\ Lett.\ {\bf 71}, 3697 (1993).
\bibitem{Murakami07} S.\ Murakami, New J.\ Phys.\ {\bf 9} 356 (2007).
\bibitem{Haldane08} F.\ D.\ M.\ Haldane and S.\ Raghu, Phys.\ Rev.\ Lett.\ {\bf 100}, 013904 (2008).
\bibitem{LLu14} L.\ Lu, J.\ D.\ Joannopoulos, and M.\ Solja\v{c}i\'{c}, Nature Photon.\ {\bf 8}, 821 (2014).
\bibitem{CHLee18} C.-H.\ Lee, {\it et al.}, Commun.\ Phys.\ {\bf 1}, 39 (2018).
\bibitem{Yatsugi22} K.\ Yatsugi, {\it et al.}, Commun.\ Phys.\ {\bf 5}, 180 (2022).
\bibitem{HYang24} H.\ Yang, L.\ Song, Y.\ Cao, and P.\ Yan, Phys.\ Rep.\ {\bf 1093}, 1 (2024).
\bibitem{Rudner20} M.\ S.\ Rudner and N.\ H.\ Lindner, Nature Rev.\ Phys.\ {\bf 2}, 229 (2020).
\bibitem{Sone24} K.\ Sone, {\it et al.}, arXiv2407.16143v1 (2024).
\bibitem{Gilbert21} M.\ J.\ Gilbert, Commun.\ Phys.\ {\bf 4}, 70 (2021).
\bibitem{HWang22} H.\ Wang, {\it et al.}, Light Sci.\ Appl.\ {\bf 11}, 292 (2022).
\bibitem{Nayak08} C.\ Nayak, {\it et al.}, Rev.\ Mod.\ Phys.\ {\bf 80}, 1083 (2008).
\bibitem{Mong14} R.\ S.\ K.\ Mong, {\it et al.}, Phys.\ Rev.\ X {\bf 4}, 011036 (2014).
%
\bibitem{XWan11} X.\ Wan, {\it et al.}, Phys.\ Rev.\ B {\bf 83}, 205101 (2011).
\bibitem{Burkov11} A.\ A.\ Burkov and L.\ Balents, Phys.\ Rev.\ Lett.\ {\bf 107}, 127205 (2011).
\bibitem{Singh12} B.\ Singh, {\it et al.}, Phys.\ Rev.\ B {\bf 86}, 115208 (2012).
\bibitem{SYXu15} S.-Y.\ Xu, {\it et al.}, Science {\bf 349}, 613 (2015).
\bibitem{SJia16} S.\ Jia, {\it et al.}, Nature Mat.\ {\bf 15}, 1140 (2016).
\bibitem{XYuan20} X.\ Yuan, {\it et al.}, Nature Commun.\ {\bf 11}, 1259 (2020).
\bibitem{NPOng21} N.\ P.\ Ong and S.\ Liang, Nature Rev.\ Phys.\ {\bf 3}, 394 (2021).
\bibitem{Burkov15} A.\ A.\ Burkov, J.\ Phys.: Condens.\ Matter.\ {\bf 27}, 113201 (2015).
\bibitem{XHuang15} X.\ Huang, {\it et al.}, Phys.\ Rev.\ X {\bf 5}, 031023 (2015).
\bibitem{Zyuzin12} A.\ A.\ Zyuzin and A.\ A.\ Burkov, Phys.\ Rev.\ B {\bf 86}, 115133 (2012).
\bibitem{CZeng22} C.\ Zeng, S.\ Nandy, and S.\ Tewari, Phys.\ Rev.\ B {\bf 105}, 125131 (2022).
%
\bibitem{Cohen18} Y.\ Cohen, {\it et al.}, PNAS {\bf 115}, 6991 (2018).
\bibitem{Draelos19} A.\ W.\ Draelos, {\it et al.}, Nano Lett.\ {\bf 19}, 1039 (2019).
\bibitem{Pankratova20} N.\ Pankratova, {\it et al.}, Phys.\ Rev.\ X {\bf 10}, 031051 (2020).
\bibitem{Graziano20} G.\ V.\ Graziano, {\it et al.}, Phys.\ Rev.\ B {\bf 101}, 054510 (2020).
\bibitem{Matsuo22NL} S.\ Matsuo, {\it et al.}, Commun.\ Phys.\ {\bf 5}, 221 (2022).
\bibitem{Graziano22} G.\ V.\ Graziano, {\it et al.}, Nature Commun.\ {\bf 13}, 5933 (2022).
\bibitem{KFHuang22} K.-F.\ Huang, {\it et al.}, Nature Commun.\ {\bf 13}, 3032 (2022).
\bibitem{Arnault22} E.\ G.\ Arnault, {\it et al.}, Nano Lett.\ {\bf 22}, 7073 (2022).
\bibitem{FZhang23} F.\ Zhang, {\it et al.}, Phys.\ Rev.\ B {\bf 107}, L140503 (2023).
\bibitem{Arnault25} E.\ G.\ Arnault, {\it et al.}, Phys.\ Rev.\ Lett.\ {\bf 134}, 067001 (2025).
\bibitem{Chiles23} J.\ Chiles, {\it et al.}, Nano Lett.\ {\bf 23}, 5257 (2023).
\bibitem{Arnault21} E.\ G.\ Arnault, {\it et al.}, Nano Lett.\ {\bf 21}, 9668 (2021).
\bibitem{Matsuo25SS} S.\ Matsuo, {\it et al.}, Phys.\ Rev.\ B {\bf 111}, 094512 (2025).
\bibitem{Matsuo23AJE} S.\ Matsuo, {\it et al.}, Sci.\ Adv.\ {\bf 9}, eadj3698 (2023).
\bibitem{Prosko24} C.\ G.\ Prosko, {\it et al.}, Phys.\ Rev.\ B {\bf 110}, 064518 (2024).
\bibitem{Matsuo23JDE} S.\ Matsuo, {\it et al.}, Nature Phys.\ {\bf 19}, 1636 (2023).
\bibitem{Gupta23} M.\ Gupta, {\it et al.}, Nature Commun.\ {\bf 14}, 3078 (2023).
\bibitem{Coraiola24JDE} M.\ Coraiola, {\it et al.}, ACS Nano {\bf 18}, 9221 (2024).
\bibitem{Matsuo23AM} S.\ Matsuo, {\it et al.}, Nature Commun.\ {\bf 14}, 8271 (2023).
\bibitem{Coraiola23AM} M.\ Coraiola, {\it et al.}, Nature Commun.\ {\bf 14}, 6784 (2023).
\bibitem{Haxell23AM} D.\ Z.\ Haxell, {\it et al.}, Nano Lett.\ {\bf 23}, 7532 (2023).
\bibitem{Strambini16} E.\ Strambini, {\it et al.}, Nature Nanotech.\ {\bf 11}, 1055 (2016).
\bibitem{Coraiola24ZE} M.\ Coraiola, {\it et al.}, Phys.\ Rev.\ X {\bf 14}, 031024 (2024).
\bibitem{Bordin23} A.\ Bordin, {\it et al.}, Phys.\ Rev.\ X {\bf 13}, 031031 (2023).
%
%
\bibitem{Yokoyama15} T.\ Yokoyama and Yu.\ V.\ Nazarov, Phys.\ Rev.\ B {\bf 92}, 155437 (2015).
\bibitem{Riwar16} R.-P.\ Riwar, {\it et al.}, Nature Commun.\ {\bf 7}, 1167 (2016).
\bibitem{Eriksson17} E.\ Eriksson, {\it et al.}, Phys.\ Rev.\ B {\bf 95}, 075417 (2017).
\bibitem{Meyer17} J.\ S.\ Meyer and M.\ Houzet, Phys.\ Rev.\ Lett.\ {\bf 119}, 136807 (2017).
\bibitem{HYXie17} H.-Y.\ Xie, M.\ G.\ Vavilov, and A.\ Levchenko, Phys.\ Rev.\ B {\bf 96}, 161406(R) (2017).
\bibitem{HYXie18} H.-Y.\ Xie, M.\ G.\ Vavilov, and A.\ Levchenko, Phys.\ Rev.\ B {\bf 97}, 035443 (2018).
\bibitem{Erdmanis18} J.\ Erdmanis, \'{A}.\ Luk\'{a}cs, and Y.\ V.\ Nazarov, Phys.\ Rev.\ B {\bf 98}, 241105(R) (2018).
\bibitem{HYXie19} H.-Y.\ Xie and A.\ Levchenko, Phys.\ Rev.\ B {\bf 99}, 094519 (2019).
\bibitem{Houzet19} M.\ Houzet and J.\ S.\ Meyer, Phys.\ Rev.\ B {\bf 100}, 014521 (2019).
\bibitem{Klees20} R.\ L.\ Klees, {\it et al.}, Phys.\ Rev.\ Lett.\ {\bf 124}, 197002 (2020).
\bibitem{Klees21} R.\ L.\ Klees, {\it et al.}, Phys.\ Rev.\ B {\bf 103}, 014516 (2021).
\bibitem{Weisbrich21} H.\ Weisbrich, {\it et al.}, Phys.\ Rev.\ X Quantum {\bf 2}, 010310 (2021).
\bibitem{YChen21A} Y.\ Chen and Y.\ V.\ Nazarov, Phys.\ Rev.\ B {\bf 103}, 045410 (2021).
\bibitem{YChen21B} Y.\ Chen and Y.\ V.\ Nazarov, Phys.\ Rev.\ B {\bf 103}, 165424 (2021).
\bibitem{Boogers22} V.\ Boogers, J.\ Erdmanis , and Y.\ V.\ Nazarov, Phys.\ Rev.\ B {\bf  105}, 235437 (2022).
\bibitem{Repin22} E.\ V.\ Repin and Y.\ V.\ Nazarov, Phys.\ Rev.\ B {\bf  105}, L041405 (2022).
\bibitem{HYXie22} H.-Y.\ Xie, J.\ Hasan, and A.\ Levchenko, Phys.\ Rev.\ B {\bf 105}, L241404 (2022).
\bibitem{Gavensky23} L.\ P.\ Gavensky, G.\ Usaj, and C.\ A.\ Balseiro, EuroPhys.\ Lett.\ {\bf 141}, 36001 (2023).
\bibitem{Septembre23} I.\ Septembre, {\it et al.}, Phys.\ Rev.\ B {\bf  107}, 165301 (2023).
\bibitem{Teshler23} L.\ Teshler, {\it et al.}, SciPost Phys.\ {\bf 15}, 214 (2023).
\bibitem{Mukhopadhyay23} A.\ Mukhopadhyay, U.\ Khanna, and S.\ Das, arXiv:2309.15159v1 (2023).
\bibitem{Matute-Canadas24} F.\ J.\ Matute-Ca\~{n}adas, L.\ Tosi, and A.\ L.\ Yeyati, Phys.\ Rev.\ X Quantum, {\bf 5}, 020340 (2024).
\bibitem{Frank24} G.\ Frank, {\it et al.}, Phys.\ Rev.\ B, {\bf 109}, 205415 (2024).
\bibitem{Zalom24} P.\ Zalom , M.\ \v{Z}onda, and T.\ Novotn\'{y}, Phys.\ Rev.\ Lett.\ {\bf 132}, 126505 (2024).
\bibitem{Ohnmacht25} D.\ C.\ Ohnmacht, {\it et al.}, Phys.\ Rev.\ Lett.\ {\bf 134}, 156601 (2025).
\bibitem{Ram25} P.\ Ram, {\it et al.}, arXiv:2501.12024v1 (2025).
\bibitem{Deb18} O.\ Deb, K.\ Sengupta, and D.\ Sen, Phys.\ Rev.\ B {\bf 97}, 174518 (2018).
\bibitem{Gavensky18} L.\ P.\ Gavensky, {\it et al.}, Phys.\ Rev.\ B {\bf 97}, 220505(R) (2018).
\bibitem{Repin19} E.\ V.\ Repin, Y.\ Chen, and Y.\ V.\ Nazarov, Phys.\ Rev.\ B {\bf 99}, 165414 (2019).
%
\bibitem{Stenger19} J.\ P.\ T.\ Stenger and D.\ Pekker, Phys.\ Rev.\ B {\bf 100}, 035420 (2019).
\bibitem{Fatemi21} V.\ Fatemi, A.\ R.\ Akhmerov, and L.\ Bretheau, Phys.\ Rev.\ Research {\bf 3}, 013288 (2021).
%
%
\bibitem{Mourik12} V.\ Mourik, {\it et al.}, Science \textbf{336}, 1003 (2012).
\bibitem{Rokhinson12} L.\ P.\ Rokhinson, X.\ Liu, and J.\ K.\ Furdyna, Nature Phys.\ {\bf 8}, 795 (2012).
\bibitem{Das12} A.\ Das, {\it et al.}, Nature Phys.\ {\bf 8}, 887 (2012).
\bibitem{Deng12} M.\ T.\ Deng, {\it et al.}, Nano Lett.\ {\bf 12}, 6414 (2012).
%
\bibitem{Zazunov17} A.\ Zazunov, {\it et al.}, Phys.\ Rev.\ B {\bf 96}, 024516 (2017).
\bibitem{Gavensky19} L.\ P.\ Gavensky, G.\ Usaj, and C.\ A.\ Balseiro, Phys.\ Rev.\ B {\bf 100}, 014514 (2019).
\bibitem{Sakurai20} K.\ Sakurai, {\it et al.}, Phys.\ Rev.\ B {\bf 101}, 174506 (2020).
\bibitem{TZhou20} T.\ Zhou, {\it et al.}, Phys.\ Rev.\ Lett.\ {\bf 124}, 137001 (2020).
\bibitem{Meyer21} J.\ S.\ Meyer and M.\ Houzet, Phys.\ Rev.\ B {\bf 103}, 174504 (2021).
\bibitem{Kenawy24} A.\ Kenawy, F.\ Hassler, and R.-P.\ Riwar, Phys.\ Rev.\ B {\bf 109}, 245423 (2024).
\bibitem{vanHeck14} B.\ van Heck, S.\ Mi, and A.\ R.\ Akhmerov, Phys.\ Rev.\ B {\bf 90}, 155450 (2014).
\bibitem{Padurariu15} C.\ Padurariu, {\it et al.}, Phys.\ Rev.\ B {\bf 92}, 205409 (2015).
\bibitem{Vischi17} F.\ Vischi, {\it et al.}, Phys.\ Rev.\ B {\bf 95}, 054504 (2017).
\bibitem{Yokoyama17} T.\ Yokoyama, {\it et al.}, Phys.\ Rev.\ B {\bf 95}, 045411 (2017).
\bibitem{Wisne24} M.\ Wisne, {\it et al.}, Phys.\ Rev.\ Lett.\ {\bf 133}, 246601 (2024).
\bibitem{Lesser21} O.\ Lesser, {\it et al.}, Phys.\ Rev.\ B {\bf 103}, L121116 (2021).
\bibitem{Lesser22} O.\ Lesser, Y.\ Oreg, and A.\ Stern, Phys.\ Rev.\ B {\bf 106}, L241405 (2022).
\bibitem{Kornich19} V.\ Kornich, H.\ S.\ Barakov, and Y.\ V.\ Nazarov, Phys.\ Rev.\ Res.\ {\bf 1}, 033004 (2019).
\bibitem{Kornich20} V.\ Kornich, H.\ S.\ Barakov, and Y.\ V.\ Nazarov, Phys.\ Rev.\ B {\bf 101}, 195430 (2020).
\bibitem{Freyn11} A.\ Freyn, {\it et al.}, Phys.\ Rev.\ Lett.\ {\bf 106}, 257005 (2011).
\bibitem{Jacquet20} R.\ Jacquet, {\it et al.}, Phys.\ Rev.\ B {\bf 102}, 064510 (2020).
\bibitem{Melin20} R.\ M\'{e}lin, Phys.\ Rev.\ B {\bf 102}, 245435 (2020).
\bibitem{Melin21} R.\ M\'{e}lin, Phys.\ Rev.\ B {\bf 104}, 075402 (2021).
\bibitem{Melin22} R.\ M\'{e}lin, Phys.\ Rev.\ B {\bf 105}, 155418 (2022).
\bibitem{Melo22} A.\ Melo, V.\ Fatemi, and A.\ R.\ Akhmerov, SciPost Phys.\ {\bf12}, 017 (2022).
\bibitem{Melin23} R.\ M\'{e}lin and D.\ Feinberg, Phys.\ Rev.\ B {\bf 107}, L161405 (2023).
\bibitem{Melin24} R.\ M\'{e}lin, C.\ B.\ Winkelmann, and R.\ Danneau, Phys.\ Rev.\ B {\bf 109}, 125406 (2024).
\bibitem{Cayao24} J.\ Cayao, P.\ Burset , and Y.\ Tanaka, Phys.\ Rev.\ B {\bf 109}, 205406 (2024).
\bibitem{Ohnmacht24} D.\ C.\ Ohnmacht, {\it et al.}, Phys.\ Rev.\ B {\bf 109}, L241407 (2024).
\bibitem{Melin16} R.\ M\'{e}lin, {\it et al.}, Phys.\ Rev.\ B {\bf 95}, 085415 (2017).
\bibitem{Nowak19} M.\ P.\ Nowak, M.\ Wimmer, and A.\ R.\ Akhmerov, Phys.\ Rev.\ B {\bf 99}, 075416 (2019).
\bibitem{Melin19} R.\ M\'{e}lin, {\it et al.}, Phys.\ Rev.\ B {\bf 100}, 035450 (2019).
\bibitem{Peyruchat21} L.\ Peyruchat, {\it et al.}, Phys.\ Rev.\ Res.\ {\bf 3}, 013289 (2021).
\bibitem{Weisbrich23} H.\ Weisbrich, {\it et al.}, Phys.\ Rev.\ Res.\ {\bf 5}, 043045 (2023).
\bibitem{Day25} I.\ A.\ Day, {\it et al.}, SciPost Phys.\ {\bf 18}, 098 (2025).
\bibitem{Alidoust12} M.\ Alidoust, G.\ Sewell, and J.\ Linder, Phys.\ Rev.\ B {\bf 85}, 144520 (2012).
\bibitem{Mai13} S.\ Mai, {\it et al.}, Phys.\ Rev.\ B {\bf 87}, 024507 (2013).
\bibitem{Moor16} A.\ Moor, A.\ F.\ Volkov, and K.\ Efetov, Phys.\ Rev.\ B {\bf 93}, 104525 (2016).
\bibitem{ZQi18} Z.\ Qi, {\it et al.}, Phys.\ Rev.\ B {\bf 97}, 134518 (2018).

\bibitem{NielsenNinomiya} H.\ B.\ Nielsen and N.\ Ninomiya, Nucl.\ Phys.\ B {\bf 185}, 20 (1981); {\it ibid.} B {\bf 193}, 173 (1981).

\bibitem{Plissard13} S.\ R.\ Plissard, {\it et al.}, Nature Nanotech.\ {\bf 8}, 859 (2013).
\bibitem{Beenakker91} C.\ W.\ J.\ Beenakker, Phys.\ Rev.\ Lett.\ {\bf 67}, 3836 (1991).
\bibitem{NazarovBlanter} Y.\ V.\ Nazarov and Y. M.\ Blanter, {\it Quantum Transport: introduction to nanoscience},
(Cambridge University Press, Cambridge, 2009).
\bibitem{Berry84} M.\ V.\ Berry, Proc.\ Royal Soc.\ A. {\bf 392}, 45 (1984).
\bibitem{Fukui05} T.\ Fukui, Y.\ Hatsugai, and H.\ Suzuki, J.\ Phys.\ Soc.\ Jpn.\ {\bf 74}, 1674 (2005).

\bibitem{com1} For Weyl semimetal, the spin-orbit interaction is essential to emerge the WPs.
Hence the energy band has a spin structure, and the sign of topological charge is related with the spin structure.
However, for the MTJJ, the spin-orbit interaction nor the spin degree of freedom is not essential.
The Andreev bound state is doublly degenerate in spin.
Then, we infer the sign of topological charge from the Chern number.

\bibitem{Zyuzin12symmetry} A.\ A.\ Zyuzin, S.\ Wu, and A.\ A.\ Burkov, Phys.\ Rev.\ B {\bf 85}, 165110 (2012).
\bibitem{Soluyanov15} A.\ A.\ Soluyanov, {\it et al.}, Nature {\bf 527}, 495 (2015).


\end{thebibliography}
\end{document}